\documentclass[letterpaper]{article} %DO NOT CHANGE THIS
\usepackage{aaai19}  %Required
\usepackage{times}  %Required
\usepackage{helvet}  %Required
\usepackage{courier}  %Required
\usepackage[hyphens]{url}  %Required
\usepackage{graphicx}  %Required
\usepackage{amsmath}
\usepackage{array}
\usepackage{booktabs}
\usepackage{floatrow}
\usepackage{graphicx}
\usepackage{hhline}
\usepackage{multirow}
\usepackage{subcaption}
\usepackage[table]{xcolor}
\usepackage{supertabular}
\usepackage{natbib}
\definecolor{lightgray}{gray}{0.9}
\usepackage{tabularx}

\newcolumntype{L}[1]{>{\raggedright\arraybackslash}p{#1}}

\begin{document}
\title{Two Computational Models for Analyzing Political Attention in Social Media}
\author{[authors removed for review]}
\author{Libby Hemphill, Angela M. Sch{\"o}pke-Gonzalez\\
School of Information\\
University of Michigan\\
Ann Arbor, MI 48104\\
\{libbyh, aschopke\}@umich.edu
}
\maketitle
\textbf{Accepted for publication in the International AAAI Conference on Web and Social Media (ICWSM 2020)}
\newline
\newline
\begin{abstract}
Understanding how political attention is divided and over what subjects is crucial for research on areas such as agenda setting, framing, and political rhetoric. Existing methods for measuring attention, such as manual labeling according to established codebooks, are expensive and can be restrictive. We describe two computational models that automatically distinguish topics in politicians' social media content. Our models---one supervised classifier and one unsupervised topic model---provide different benefits. The supervised classifier reduces the labor required to classify content according to pre-determined topic list. However, tweets do more than communicate policy positions. Our unsupervised model uncovers both political topics and other Twitter uses (e.g., constituent service). These models are effective, inexpensive computational tools for political communication and social media research. We demonstrate their utility and discuss the different analyses they afford by applying both models to the tweets posted by members of the 115th U.S. Congress.
\end{abstract}

\noindent Questions about which political topics receive attention and how that attention is distributed are central to issues such as agenda setting and framing. Knowing what politicians are talking about and how those topics differ among various populations (e.g., Democrats and Republicans) and over time could enable advances in political communication research and has potential to increase constituents' knowledge. As social media becomes an increasingly common site of political discussion and impact, it becomes possible to exploit the data social media generates to understand political attention \citep{Barbera2018-bb,Neuman2014-tb}. Twitter activity, especially, relates to media and elites' political attention \citep{Shapiro2017-ur,Guggenheim2015-bk, Rill2014-mx} and offers an opportunity to measure attention and its changes over time.

We use data from Twitter to address two primary challenges in studying political attention and congressional communication. First, existing methods for studying political attention, such as manual topic labeling, are expensive and restrictive. Second, our methodological tools for studying Congress on Twitter have not kept pace with Congress's adoption and use. 

To address these challenges, we developed two computational models for estimating political attention---one supervised model that leverages human labels to classify texts at scale and a second unsupervised model that uses readily-available data and low computational overhead to automatically classify social media posts according to their policy topic and communication style. By providing the models and their codebooks, we enable others to label political texts efficiently and automatically. This facilitates efficient and nuanced studies of Congress's political attention and communication style and supports comparative studies of political attention across media (i.e., social media, congressional hearings, news media). We demonstrate the utility of these approaches by applying the models to the complete corpus of tweets posted by the 115th U.S. Congress and briefly discussing the insights gained. We also discuss the costs and benefits of supervised and unsupervised models and provide aspects of each to consider when choosing a tool for analysis. In summary, our contributions are

\begin{enumerate}
    \item a supervised model for assigning tweets to a pre-existing set of policy topics and facilitating comparative analyses;
    \item an unsupervised model for assigning tweets to inferred policy topics and communication uses; and
    \item trade offs to consider when choosing supervised and/or unsupervised approaches to topic labeling.
\end{enumerate}

Labeling content by hand requires tremendous human effort and substantial domain knowledge \citep{Quinn2010-nt}. Attempts to crowd source annotation of political texts acknowledge that domain expertise is required for classifying content according to existing policy code books \citep{Benoit2016-ik,Haselmayer2017-dj,Lehmann2018-om}. Using pre-defined codebooks assumes a particular set of topics and therefore cannot effectively classify content that falls outside those predefined areas; however, codebooks enable comparisons across governing bodies, over time, and among groups. Quinn and colleagues~\citeyearpar{Quinn2010-nt} provide a more detailed overview of the challenges associated with labeling according to known topics and by hand; they also provide a topic-modeling approach to classifying speech in the Congressional Record that is similar to ours. 

Our approach differs from Quinn et al.'s in the types of documents we use for the model (speeches from the Congressional Record vs tweets) and the model's assumptions. They employ a dynamic model that includes time parameters that distinguish days in session from days not in session, and our approach uses a static latent Dirichlet model (LDA) similar to Barber\'a et al.'s \citeyearpar{Barbera2018-bb,Barbera2014-pa}. The static model is appropriate in our case because tweets do not exhibit the time parameters of Congressional speeches (i.e., speeches at time \textit{t} and \textit{t} + 1 are likely related while the same is not necessarily true for tweets). Our approach differs from Barber\'a et al.'s in that we use individual tweets rather than aggregates by day, party, and chamber. Despite earlier work that suggests tweets are too short for good topic models \citep{Hong2010-uh}, we found acceptable performance and useful sensitivity by using individual tweets \citep[see ][]{Zhao2011-me}.

Congress increasingly uses social media as a mechanism for speaking about and engaging with constituencies~\citep{Straus2013-mb}, but our tools for studying their social media use have not kept pace. Existing studies of members of Congress's (MCs') social media use rely on the human labeling and domain knowledge mentioned above~\citep{Russell2017-ee,Russell2018-vu,Evans2014-or,Frechette2017-sq} or focus on the frequency~\citep{LaMarre2013-zr} or type of use~\citep{Golbeck2018-py,Hemphill2013-rb} rather than the content of messages. MCs' social media speech can be used for understanding polarization \citep{Hemphill2016-fc,Hong2016-xq} and likely impacts political news coverage \citep{Shapiro2017-ur,Moon2014-xl}, and new methodological tools for estimating attention and style would provide richer views of activity and enable new analyses.

\subsubsection{Why Twitter} 
Existing work in political attention relies largely on political speeches \citep{Laver2003-id,Oliver2016-pz,Yu2008-wq,Quinn2010-nt} and party manifestos \citep{Gabel2000-gh,Slapin2008-kk,Benoit2016-ik}. At the same time, politicians around the world increasingly use social media to communicate, and researchers are examining the impacts of that use on elections \citep{Bossetta2018-yc,Karlsen2011-hs}, the press \citep{Murthy2015-mg,Shapiro2017-ur}, and public opinion \citep{Michael2018-uk}. Given its prevalence among politicians and in the public conversation about politics, politicians' behavior on Twitter deserves our attention. We also expect that Twitter data's availability and frequent updating will enable us to study political attention more efficiently and at a larger scale than prior data sources. 

In order to understand how U.S. Congressional tweets may help us to answer social and political research questions (e.g., How does MC attention change over time? Does social media influence the political agenda? Does Congress respond to the public's policy priorities?), we must first understand what U.S. MCs are saying. Given the volume of content that MCs generate and the expertise required to classify it, manual labeling is neither efficient nor affordable in the long term. We therefore explore whether computational labeling approaches could help us understand MC conversations on Twitter.

Computational study of MCs use of Twitter will eventually allow us to (1) situate the MCs' political attention patterns on Twitter within the broader media context, (2) understand whether MCs' political attention patterns on Twitter reflect attention patterns in a broader media context, and (3) understand the unique value that the use of supervised classification and unsupervised topic modeling techniques can contribute to our overall understanding of MCs' political attention patterns. % However, though approaches to studying attention computationally have been described \citep{Quinn2010-nt,Barbera2018-bb,Barbera2014-pa,Benoit2016-ik}, tools or models for performing these analysis are currently unavailable.

\section{Data}

\subsection{Tweets from the 115th U.S. Congress}

Using the Twitter Search API, we collected all tweets posted by official MC accounts (voting members only) during the 115th U.S. Congress which ran January 3, 2017 to January 3, 2019. We identified MCs' Twitter user names by combining the lists of MC social media accounts from the United States project\footnote{\url{https://github.com/unitedstates/congress-legislators}}, George Washington Libraries\footnote{\url{https://dataverse.harvard.edu/dataset.xhtml?persistentId=doi:10.7910/DVN/UIVHQR}}, and the Sunlight Foundation\footnote{\url{https://sunlightlabs.github.io/congress/index.html\#legislator-spreadsheet}}. 

\begin{table}
\begin{tabular}{lrrrrr}
\toprule
  & \multicolumn{2}{c}{Senate} & \multicolumn{2}{c}{House} & Total \\
\hline
 & Dem & GOP & Dem & GOP & \\
\hline
 Men & 32 & 48 & 128 & 207 &  415\\
 Women & 16 & 5 & 64 & 24 & 109 \\
 Total & 48 & 53 & 192 & 231 & 524 \\
\bottomrule
\end{tabular}
\caption{Number of accounts by party, chamber, and gender. Two independent senators who caucus with Democrats have been grouped with them.}\label{table:accounts}
\end{table}

\begin{table}
\begin{tabular}{lrrrr}
\toprule
   & \multicolumn{2}{c}{Senate} & \multicolumn{2}{c}{House}\\
\hline
 & Dem & GOP & Dem & GOP \\
\hline
 Men & 143,522 & 171,083 & 441,890 & 356,870 \\
 Women & 74,905 & 19,867 & 212,457 & 65,240 \\
 Total & 218,427 & 190,950 & 654,347 & 422,110\\
\bottomrule
\end{tabular}
\caption{Number of tweets by party, chamber, and gender. Two independent senators who caucus with Democrats have been grouped with them. Cell values are the count of tweets posted by the intersection of the row and column. Men posted 1,113,365 tweets, and women posted 372,469.}\label{table:frequencies}
\end{table}

Throughout 2017 and 2018, we  used the Twitter API to search for the user names in this composite list and retrieved the accounts' most recent tweets. Our final search occurred on January 3, 2019, shortly after the 115th U.S. Congress ended. In all, we collected 1,485,834 original tweets (i.e., we excluded retweets) from 524 accounts. The accounts differ from the total size of Congress because we included tweet data for MCs who resigned (e.g., Ryan Zinke) and those who joined off cycle (e.g., Rep. Conor Lamb); we were also unable to confirm accounts for every state and district. We summarize the accounts by party, chamber and gender in Table \ref{table:accounts} and the number of tweets posted by chamber, gender, and party in Table \ref{table:frequencies}. % We used this data to train our unsupervised topic model, and to compare our supervised model's performance against our unsupervised model's performance.

\section{Two Models for Classifying Tweets According to Political Topics}

\subsection{Preprocessing}
We used the same preprocessing steps for both the supervised and unsupervised models:

\subsubsection{Stemming}

We evaluated whether or not to use stemming or lemmatization in training our models by reviewing the relative interpretability of topics generated by models trained with each of stemmed texts and unstemmed texts. We found unstemmed texts render significantly more interpretable generated topics, likely reflecting syntactical associations with semantic meanings, which is consistent with prior work~\citep{Schofield2016-xp}. This pattern is likely especially relevant for tweet data given that unique linguistic patterns and intentional misspellings are used to communicate different semantic meanings. Stemming in these instances may remove the nuance in potential semantic meaning achieved by tweets' unique linguistic features including misspellings. Therefore, we did not use stemming or lemmatization in data preprocessing for the final models.

\subsubsection{Stop lists, tokenization, and n-grams}

Given the prevalence of both English- and Spanish-language tweets, this project removed both English and Spanish stop words included in Python's Natural Language Toolkit (NLTK) default stop word lists \citep{Bird2009-vi}. 

We also used a combination of two tokenization approaches to prepare our data for modeling. First, we used Python's NLTK tweetTokenizer with parameters set to render all text lower case, strip Twitter user name handles, and replace repeated character sequences of length three or greater with sequences of length three (i.e. ``Heeeeeeeey" and ``Heeeey" both become ``Heeey"). Second, we removed punctuation (including emojis), URLs, words smaller than two letters, and words that contain numbers.

In the early stages of model development, we evaluated the comparative intepretability and relevance of topics generated for document sets tokenized into unigrams, bigrams, and a combination of unigrams and bigrams. We found model results for document sets tokenized into unigrams to be most interpretable and relevant, and thus present only these results here. 

\subsection{Specifying Models}

\subsubsection{Document Selection}
We used the same documents for each model: individual, original tweets. We compared the coherence and interpretability of models using other document types (e.g., author-day, author-week, and author-month aggregations in alignment with Quinn et al. \citeyearpar{Quinn2010-nt} and Barberá et al.'s \citeyearpar{Barbera2018-bb} findings that aggregations improve model performance) and found models using individual tweets to produce interpretable, facially valid topics.

We expect that the individual tweet approach works well because individuals are discussing more multiple topics per day. Congressional speeches, like those Quinn et al. use, are likely more constrained than tweets over short periods, and therefore are appropriate to aggregate. Aggregations over long periods such as author-week and author-month documents produced even less distinct topics.

\subsubsection{Model Output}

Both models produce the same type of output. The models determine the probability that a given tweet belongs in each of the topics (\textit{categories} for the supervised model and \textit{inferred topics} for the unsupervised) and then assigns the highest-probability category. We also report the second-most likely category assigned by the unsupervised model in cases where mutually exclusive categories are not required or where a second class provides additional information about the style or goal of the tweets. These outputs mean that the supervised model's predictions are directly aligned with the CAP codebook, and the unsupervised model's predictions capitalize on its ability to relax requirements and to discover multiple topics and features within tweets.

\subsubsection{Supervised Classifier}

We trained a supervised machine learning classifier using labeled tweets from Russell's research on the Senate \citeyearpar{Russell2017-ee,Russell2018-vu}. The dataset  contains 68,398 tweets total: 45,402 tweets labeled with codes from the Comparative Agenda Project's (CAP's) codebook \citep{Bevan2017-yw} and 22,996 labeled as not-policy tweets. The CAP codebook is commonly used in social, political, and communication science to understand topics in different types of political discourse worldwide \citep[see, e.g., ][for the project's introduction and recent collections of research]{Baumgartner2019-sx,Baumgartner1993-qq,John2006-rx} and to evaluate unsupervised topic modeling approaches \citep{Quinn2010-nt, Barbera2018-bb}. Recent discussions of the CAP codebook center around its mutually exhaustive categories and backward compatibility when discussing its use as a measurement tool in political attention research \citep{Jones2016-hi, Dowding2016-sn}. We removed retweets to limit our classification to original tweets, resulting in a total set of 59,826 labeled tweets (39,704 policy tweets and 20,122 not policy tweets). The training set was imbalanced, and we found using a subset of the over-represented not-policy tweets in training improved the model's performance. Our final training corpus included 41,716 tweets (39,704 policy tweets and 2,012 not policy tweets).

We trained and tested four types of supervised classification models using NLTK \citep{Bird2009-vi} implementations: a random guessing baseline dummy model (D) using stratified samples that respect the training data's class distribution, a Na\"ive Bayes (NB) model, a Logistic Regression (LR) model, and a Support Vector Machine (SVM) model. In each case, we used a 90-10 split for train-test data meaning that 90\% of labeled tweets were used as training instances, and the models then predicted labels for the remaining 10\%. 
%We then compared the models’ predictions with the human labels to evaluate their performance. 
After initial testing, for each of our top-performing models (LR and SVM), we evaluated whether the addition of Word2Vec \citep[W2V; ][]{Mikolov2013-ja} word embedding features or Linguistic Inquiry and Word Count \citep[LIWC; ][]{Pennebaker2015-iq} features could improve their performance. We extracted all LIWC 2015 features for each token in our Tweet token corpus and integrated them as features into our model. In all cases, we used a 90/10 split for training/test data.

\subsubsection{Unsupervised Topic Model}
We present the results of the unsupervised model generated using MALLET's LDA model wrapper \citep{rehurek_lrec}. We tested other models (described in more detail below) but found the MALLET LDA wrapper produced the most analytically useful (i.e., interpretable, relevant) topics. The MALLET LDA wrapper requires that we specify a number of topics to find in advance. We evaluated the performance of models generating between 5 and 70 topics in increments of five topics within this range. We found 50 topics to yield the most interpretable, relevant, and distinct results.

We found that the results of the gensim LDA model were insufficiently interpretable to provide analytic utility. We also considered Moody's lda2vec \citeyearpar{moody2016}, and though preliminary results did return more nuanced topics, we found its setup and computational inefficiency to be too cumbersome and costly relative to the benefit of that nuance. LDA models are a popular and effective choice in recent topic modeling work, though much of that work uses supervised approaches \citep{Mcauliffe2008-aa, Nguyen2013-ln, Resnik2015-aw, Perotte2011-ff}. Therefore, we considered Supervised LDA \citep[sLDA][]{Mcauliffe2008-aa} as a third alternative approach, but found it even less computationally efficient than lda2vec. Though they often demonstrate improved performance over LDA models, we did not use Structural Topic Modeling (STM) techniques \citep{RobertsME2014-stm} because of their assumptions about the relationships between metadata (e.g., party affiliation) and behavior are actually the objects of study in our use case---we cannot study party differences if we include party in the model's specifications.

\subsection{Evaluating Models}

% \subsubsection{Evaluating the Supervised Model}
We summarize the performance of our supervised models in Table \ref{table:supervised-models}; the models performed quite similarly despite variations in their algorithms and features. Our highest-performing supervised classifier (logistic regression) achieved an F1 score of 0.79. The F1 score balances the precision and recall of the classifier to provide a measure of performance. Given the difference between our classifier's score and the dummy, we argue the supervised classifier achieved reasonable accuracy. % We also compared our classifier's labels with the human labels and achieved a Cohen's kappa of 0.78, suggesting moderate agreement \citep{McHugh2012-yb}.

% \subsubsection{Evaluating the Unsupervised Model}

We used an inductive approach to interpret and label topics returned by our final unsupervised model. We first labeled the baseline model's topics according to our initial interpretations, allowing us to discover semantically-important topics that arise from the data rather than from a predetermined topic set. Human interpretation and label assignment was performed by two domain experts and confirmed by a third. One expert has prior experience on legislative staff having served in a senior senator's office and in public affairs for political organizations. She was able to determine whether the topics identified by the classifier were interpretable and were actually capturing political topics as members of Congress understand them. The second expert has spent 12 years working in political communication research, allowing her to understand topics returned within the context of political behavior. The two labelers discussed disagreements until both agreed on the labels applied to all fifty topics. Following this first labeling process, we looked for patterns across labels and grouped labels together that indicated topical similarity based on their highest-weighted features. For instance, one topic's top unigram features included \textit{health, care, Americans, Trumpcare} and another included \textit{health, care, access, women}. We bundled both of these topics under the umbrella \textit{healthcare}. Finally, we reviewed our topic labels and features with a senior member of a U.S. policy think tank to confirm the validity of our labels and interpretability of our topics.

\begin{table}[th]
\begin{tabular}{llll}
\toprule
Classifier                     & F1 & Prec. & Recall\\
\midrule
D                              & 0.07     & 0.07    & 0.07             \\
NB                             & 0.71     & 0.74    & 0.72          \\
\textbf{LR}                    & \textbf{0.79}     & \textbf{0.79}    & \textbf{0.79}          \\
LR + pre-trained w2v features  & 0.77     & 0.78    & 0.77          \\
LR + original w2v features     & 0.78     & 0.79    & 0.78          \\
LR + LIWC                      & 0.78       & 0.78  & 0.78          \\
% & 0.78     & 0.76      
SVM                            & 0.78     & 0.79    & 0.78          \\
SVM + pre-trained w2v features & 0.77     & 0.78    & 0.77          \\
SVM + original w2v features    & 0.78     & 0.79    & 0.77          \\
SVM + LIWC                     & 0.77     & 0.78  & 0.77\\  
% 0.78     & 0.76
\bottomrule
\caption{Supervised Classifier Performance Summary: dummy (D), Na\"ive Bayes (NB), Logistic Regression (LR) model, and Support Vector Machine (SVM) models are included. Features included unigrams, word vectors (w2v) and LIWC features}\label{table:supervised-models}
\end{tabular}
\end{table}

\begin{table}[ht!]
\begin{tabular}{lrrrr}
\toprule
Topic Description & \# & SU & UN1 & UN2\\
\midrule
macroeconomics &1           &  0.060 &  0.074 &  0.042 \\
civil rights & 2           &  0.037 &  0.094 &  0.040 \\
health & 3           &  0.088 &  0.318 &  0.043 \\
agriculture & 4           &  0.010 &  0.019 &  0.014 \\
labor & 5           &  0.022 &  0.021 &  0.016 \\
education & 6           &  0.021 &  0.021 &  0.016 \\
environment & 7           &  0.018 &  0.045 &  0.012 \\
energy & 8           &  0.013 &       - &       - \\
immigration & 9           &  0.021 &  0.017 &  0.012 \\
transportation & 10          &  0.011 &       - &  0.020 \\
law and crime & 12          &  0.043 &  0.026 &  0.018 \\
social welfare & 13          &  0.009 &  0.020 &  0.014 \\
housing & 14          &  0.003 &       - &       - \\
domestic commerce & 15          &  0.027 &  0.023 &  0.018 \\
defense & 16          &  0.066 &       - &  0.019 \\
technology & 17          &  0.006 &       - &  0.005 \\
foreign trade & 18          &  0.002 &       - &       - \\
international affairs & 19          &  0.027 &       - &       - \\
government operations & 20          &  0.037 &  0.072 &  0.059 \\
public lands & 21          &  0.009 &       - &  0.010 \\
cultural affairs &	23 &   - &   - &  -\\
veterans & 24          &       - &  0.019 &  0.026 \\
sports & 25          &       - &  0.020 &  0.011 \\
district affairs & 26          &       - &  0.071 &  0.063 \\
holidays & 27          &       - &  0.009 &  0.027 \\
awards & 28          &       - &       - &  0.009 \\
politicking & 29          &       - &       - &  0.042 \\
self promotion & 30          &       - &  0.050 &  0.052 \\
sympathy & 31          &       - &       - &  0.010 \\
emergency response & 32          &       - &  0.018 &  0.013 \\
legislative process & 33          &       - &       - &  0.059 \\
constituent relations & 34          &       - &       - &  0.022 \\
power relations & 35          &       - &  0.026 &  0.024 \\
uninterpretable & -           &  - &  0.037 &  0.284 \\
\bottomrule
\caption{Topic Distribution Across the 115th U.S. Congress. Contains proportional distributions across all topics for both supervised (SU) and unsupervised (UN1 and UN2) classifiers. UN1 indicates the highest probability category assigned, and UN2 indicates the second highest. Topics 1-23 correspond to labels in the CAP codebook, and 24-35 are new codes identified by our unsupervised model.}\label{table:topic-distribution}
\end{tabular}
\end{table}

Where possible, we matched CAP codes to the related codes that resulted from our inductive labeling. A complete list of our topics, their associated CAP codes, and their frequencies is available in Table \ref{table:topic-distribution}.\footnote{The CAP codebook does not include codes 11 or 22.} Not all of our codes had ready analogues in the CAP taxonomy. For instance, our unsupervised model identified \textit{media appearance} as a separate and frequent topic. The CAP codebook includes only policy areas and not relationship-building or constituent service activities, and so no CAP label directly applied there. Further, we determined that some nuanced aspects of the topics returned by our unsupervised model are not captured by the policy focus of the CAP codebook. For example, our unsupervised model returned several topics clearly interpretable as related to veteran affairs. The CAP codebook includes aspects of veteran affairs as a sub-topic under both \textit{housing} and \textit{defense}, but given that our unsupervised model detected veteran-related issues outside of these sup-topic areas, we determined that a new high-level topic devoted solely to \textit{veteran affairs} would better describe the thematic content of those topics. A similar situation applied to the topic we labeled \textit{legislative process}, some but not all aspects of which may have fallen under CAP code \textit{government operations}. The codes we identified that were not captured by the CAP codebook are marked by ``-" in the Code Number column of Table \ref{table:topic-distribution}.

\subsubsection{Topic Coherence}
Topic coherence---numerical evaluation of how a ``fact set can be interpreted in a context that covers all or most of the facts'' present in the set \citep{TopicEval2015}---was one consideration when deciding whether we had found the right number and mix of topics using the unsupervised classifier. Recent work has sought to quantitatively evaluate topic model performance using measures such as perplexity \citep{Barbera2018-bb}. However, perplexity actually correlates negatively with human interpretability \citep{NIPS2009_3700}. As an alternative to perplexity, Lau et al. \citeyearpar{Lau2014-qp,Lau2016TheSO} and Fang et al. \citeyearpar{Fang2016-cf} provide topic coherence evaluation measures that use point wise mutual information (NPMI and PMI) and find them to emulate human interpretability well.

Given this existing research, we implemented the NPMI topic coherence measure \citep{Lau2014-qp,Lau2016TheSO} to evaluate early iterations of our topic models.
Reference corpus selection is an especially important component of ensuring that the NPMI topic coherence metric provides a useful measure by which to evaluate a topic models performance. %Fang et. al posit that Tweets unique syntax require that a reference corpus reflects this syntactical difference. As such, they used an additional selection of random Tweets from a given date range as their reference corpus \citep{Fang2016-cf}. In this project’s context, however, we hypothesized that a selection of random tweets would not capture the domain-specific context unique to Congressional Members Tweets. As such, 
We tested two reference corpora: a set of random tweets and all U.S. legislative bill text from the 115th U.S. Congress.

Initial results yielded very low topic coherence values across both reference corpora in relation to those topic coherence measures reported elsewhere \citep{Lau2014-qp}. This discrepancy likely indicates that random tweets are insufficiently specific and that bill text is too syntactically dissimilar from MC tweets for either corpus to be an appropriate reference text. Therefore, for final model iterations we evaluate our models solely according to human interpretability. %insufficient domain-specific content in the random tweet sample as hypothesized, and perhaps that the bill text reference corpus does not reflect sufficient syntactic similarity to Twitter data to be useful and make up for insufficient domain-specific content in the random tweet reference corpus. Further analysis of these results indicated that highest topic coherence scores resulted for inconsistent document types. Each of three document types, individual tweets, author-week, and author-month, resulted in the highest topic coherence score for each of 15, 20, and 25 topics tested respectively. Of these highest scores, the highest was returned by 15 topics with author-week documents. However, in our human interpretability evaluations, we found that individual tweet documents with 50 topics provided the most meaningful returned topics. 

% These results taken into consideration, we found that higher topic coherence outputs did not correlate with human interpretability in any meaningful way. As such, 

\subsection{Comparing Models' Output}
The supervised model provides a single topic label for each tweet, and the unsupervised model provides probabilities for each tweet-class pair. In order to compare the labels between classifiers, we assigned tweets the highest-probability topic indicated by the unsupervised model. We measured interannotator agreement between our supervised and unsupervised models using Cohen's kappa \citep{McHugh2012-yb}. We opted to use Cohen's kappa as a measure that is widely-used for comparing two annotators \citep{Hellgren2012}. Because our supervised model assigns topic labels only to policy-related tweets, a comparison between supervised and unsupervised models is only meaningful for those tweets both models labeled, or policy-related tweets. Therefore, we calculated a Cohen's kappa score between our supervised and unsupervised model for only policy-related tweets with label assignments corresponding to CAP codes. 

The two classifiers achieved a Cohen's kappa of 0.262. We argue that the classifiers likely measure different things, and the low agreement suggests an opportunity for researchers to select the tool that matches their analytic goals. Low agreement suggests that the two models are well-differentiated; we return to this discussion in detail below. The features associated with each models' classes are presented in tables \ref{table:su-features} and \ref{table:un-features} and also indicate that the models are distinct.

\begin{table}[ht!]
\begin{tabularx}{\linewidth}{l X}
\toprule
Label & Associated Terms \\
\midrule
macroeconomics           &  budgetconference fiscalcliff budgetdeal \\
civil rights           &  passenda enda nsa  \\
health           &  defundobamacare obamacare healthcare  \\
agriculture           &  farmbill gmo sugar  \\
labor           &  fmla minimumwage laborday  \\
education           &  talkhighered dontdoublemyrate studentloans  \\
environment           &  actonclimate leahysummit climate \\
energy           &  energyefficiency energyindependence \\
immigration           &  immigrationreform immigration cirmarkup \\
transportation          &  skagitbridge obamaflightdelays faa \\
law and crime          &  vawa guncontrol gunviolence  \\
social welfare          &  snap nutrition hungry \\
housing          &  gsereform fha housing  \\
domestic commerce          &  fema sandyrelief sandy  \\
defense          &  veteransday drones stolenvalor  \\
technology          &  marketplacefairness nonettax broadband  \\
foreign trade          &  trade exports export  \\
international affairs          &  benghazi standwithisrael  \\
government operations          &  nomination irs nominations \\
public lands         &  commissiononnativechildren tribalnations \\
\bottomrule
\end{tabularx}
\caption{Features Associated with Topics in the Supervised Classifier}\label{table:su-features}
\end{table}

\begin{table}[ht!]
\begin{tabularx}{\linewidth}{lX}
\toprule
Label & Associated Terms \\
\midrule
macroeconomics           &  budget government house bill; tax families cuts; tax jobs reform \\
civil rights           &  trump president american today life rights women rights equal \\
health           &  health care americans; opioid help health; health getcovered enrollment  \\
agriculture           &  energy jobs farmers \\
labor           &  obamacare jobs year \\
education           &  students education student  \\
environment           &  climate change water  \\
immigration           &  children border families \\
transportation          &  will infrastructure funding  \\
law and crime          &  sexual human trafficking; gun violence congress \\
social welfare          &  families workers care  \\
domestic commerce          &  small jobs businesses \\
defense          &  iran nuclear security  \\
technology          &  internet netneutrality open \\
government operations          &  act bill protect; court judge senate; trump investigation russia; obama rule president  \\
public lands          &  national protect public  \\
veterans          &  veterans service honor; women service men \\
sports          &  game team win \\
district affairs          &  office help hours; today great new; service academy fair; hall town meeting \\
holidays          &  day happy today; family will friend; happy year celebrating \\
awards          &  school congressional art \\
politicking          &  make work keep; forward work working  \\
self promotion          &  read week facebook; tune watch live; hearing watch committee  \\
sympathy          &  families victims prayers  \\
emergency response          &  help disaster hurricane  \\
legislative process          &  bill house act; today discuss issues; vote voting voter \\
constituent relations          &  work community thank  \\
power relations          &  today president mayor  \\
\bottomrule
\end{tabularx}
\caption{Features Associated with Topics in the Unsupervised Model. Longer feature lists indicate that multiple unsupervised topics were merged into a single parent topic.}\label{table:un-features}
\end{table}

\section{Applying Models to 115th U.S. Congress's Tweets}
We applied both the supervised and unsupervised models to all of the original tweets posted by the 115th U.S. Congress and use those results to illustrate the analytic potential of these new methodological tools. % Table \ref{table:topic-distribution} summarizes the results by providing the topic description, the corresponding CAP codebook number (if applicable), the proportion of tweets that fell in that topic according to the supervised (SU) and unsupervised (UN1 and UN2) models. For each tweet, the unsupervised model returns the probability that the tweet belongs in each of the 50 topics identified. We provide both the most likely (UN1) and second-most likely topic (UN2) for each tweet.

\subsubsection{Topic Distribution}

In Table \ref{table:topic-distribution}, each of topics 1-23 correspond to the CAP macro-level code numbers. Topic 23, \textit{cultural affairs} was not detected by either of our models, and as such receives a ``-'' value across all columns. Each topic 24-35 corresponds to expert interpretations of unsupervised topic model results; we assigned ``0'' to tweets whose topics were uninterpretable by either model. Since these latter topics apply only to the unsupervised model's results, each of these topics receives a ``-'' value in column ``SU''. Related, each of topics \textit{energy}, \textit{housing}, \textit{foreign trade}, and \textit{international affairs} received ``-'' values in  column ``UN1''). These ``-'' values indicate that our unsupervised model did not detect topics that human interpretation would assign one of these four CAP codebook topics. 

\begin{figure*}[h!]
\centering
    \subcaptionbox{}{    \includegraphics[width=0.3\textwidth]{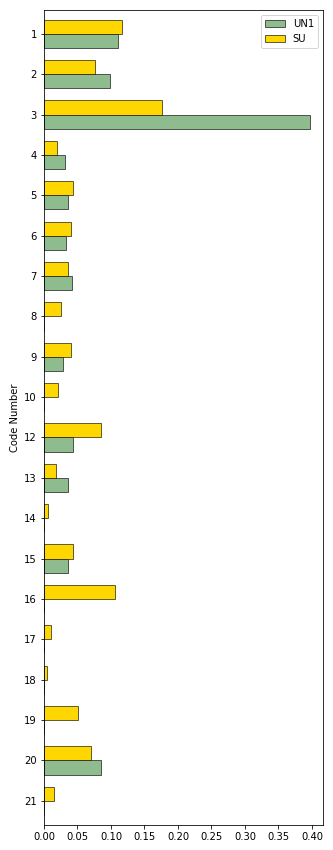}}
    \subcaptionbox{}{    \includegraphics[width=.3\textwidth]{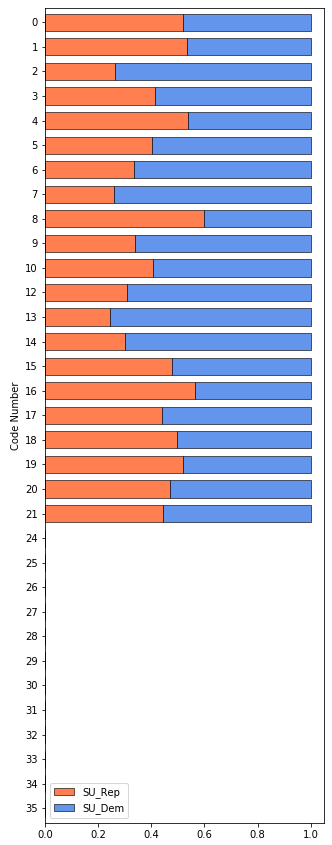}}
    \subcaptionbox{}{    \includegraphics[width=.3\textwidth]{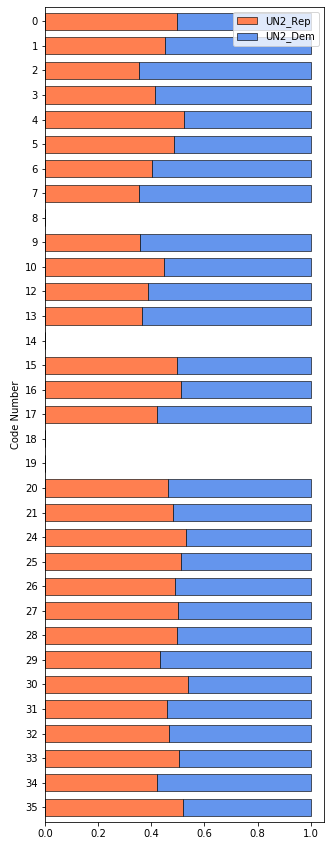}}
\caption{Topic Distributions. (a) shows the overall topic distribution according to the supervised classifier. (b) contains a 100\% stacked bar chart indicating the proportion of tweets of that type that came from Republicans (red) and Democrats (blue) as labeled by the supervised classifier. (c) shows the proportion of tweets of that type that came from Republicans (red) and Democrats (blue) as indicated by the second-highest topic according to the unsupervised classifier.}
\label{figure:topic-distributions}
\end{figure*}

\section{Discussion}
Examining the output from the two models on the same dataset, tweets from the 115th U.S. Congress, helps explain the models' differences, trade-offs, and clarifies the new insights available from the unsupervised model.

\subsection{Comparing Model Labels} \label{comparing-models}

First, we examined the proportion of supervised labels matching most likely unsupervised labels for policy-related tweets. Table \ref{table:topic-distribution} shows that both models were able to detect a topic in nearly all tweets (i.e., the proportion of \textit{uninterpretable} appears relatively infrequently). %This topic appears more frequently in the second set of labels from the unsupervised classifier (UN2). 
The \textit{uninterpretable} topic's features were comprised largely of prepositions, conjunctions, and other words with little semantic meaning. This suggests that the most significant semantic meaning of a tweet labeled by the unsupervised model can likely be understood by using its maximum probability topic.

In Figure \ref{figure:topic-distributions} (a), we include the supervised model and most-likely topic from the unsupervised model because those two topics are policy-related. In Figure \ref{figure:topic-distributions} (b) and (c), we report the supervised classifier label and second-most likely unsupervised topic label to illustrate how it's possible to use each classifier for different analytic purposes. The supervised classifier, Figure \ref{figure:topic-distributions} (b), shows us differences in political attention by party. The unsupervised classifier's second-most likely, see Figure \ref{figure:topic-distributions} (c), category enables us to compare the style by party. We can see that Republicans give less attention to \textit{civil rights}, \textit{environment}, and \textit{social welfare} and more attention to \textit{energy} and \textit{defense} than Democrats. None of these differences are especially surprising given the parties' priorities, but our models show empirically that the differences are observable even in social media. Among the unsupervised model's style codes, Republicans exhibit more \textit{self promotion} than Democrats, but the parties are nearly equal on those style codes. Our model affords similar comparisons among other groups such as gender or chamber that may provide insight into how political attention changes or how communication styles differ among groups.

Additional similarities and differences in the proportions of SU and UN2 in Figure \ref{figure:topic-distributions} may help us to understand the utility of each model.  We can see that our unsupervised model's maximum probability topic predictions did not include topic labels \textit{transportation}, \textit{defense}, 
\textit{technology}, \textit{public lands} for any policy-related tweets. However, we can also see that each of these topics are included among the unsupervised model's second most probable topic predictions. This suggests that in some policy topics' cases, each of our supervised and unsupervised models may be able to predict approximately similar ideas if both most probable and second most probable topics are taken into account.

Interestingly, we see that the \textit{government operations} topic is predicted with about the same frequency by the supervised model and each of the unsupervised model's first most probable topic assignment and second most probable topic assignment. This again suggests that the unsupervised model and supervised model both are able to predict approximately similar ideas. %, but also begs the question, which was the first most probable topic assignment for each of those tweets that were labeled \textit{government operations} with second-highest probability?

In general, we notice that among the second-most probable topics for the unsupervised model, topics 24-35, or all those topics that are not featured in the CAP codebook, occur more frequently. This suggests that the topic of highest probability more effectively captures a policy focus, and the topic of second-highest probability captures the way in which, the reason for, and with whom MCs are discussing these policy issues.

\subsection{Rare Topics}
The unsupervised model did not detect the following CAP codebook topics: \textit{energy}, \textit{housing}, \textit{foreign trade}, \textit{international affairs}. This likely indicates that these topics are occurring rarely enough that the unsupervised model does not have enough data to detect these topics as distinct from others. In reviewing Figure \ref{figure:topic-distributions}, it is possible to see that a very low proportion of tweets were labeled \textit{housing} or \textit{foreign trade} by the supervised classifier, for example. That these topics are not frequently occurring according to the supervised model either suggests that that MCs do not spend a lot of time tweeting about those topics. These omissions are likely an artifact of the U.S. federal government's structure. Though housing, for example, is in part an issue addressed by the U.S. Department of Housing and Urban Development and thus an appropriations issue for the U.S. Congress, many housing issues are addressed at state and local levels. It is possible that given this distribution of significant responsibility to the state and local levels, MCs spend less time talking about housing at the federal level. 

% \begin{table}[t!]
% \begin{tabular}{lrrrr}
% \toprule
% Model 1 &  Model 2 &  Cohen's Kappa Score\\
% \midrule
% Unsupervised &    Supervised &    0.012\\
% Unsupervised &    Hand-Labeled &    0.013\\
% % Supervised &    Hand-Labeled &    0.960\\
% \bottomrule
% \caption{Comparing Model Output to Manual Labels}\label{table:model-manual}
% \end{tabular}
% \end{table}

\subsection{Where the Models Diverge}

It is also possible that those terms associated with these topics for the supervised model are being associated with or grouped together with terms corresponding to different topics. For example, given that \textit{international affairs}, \textit{defense}, and \textit{immigration} topic areas have some overlapping topical relevance, it is possible that certain tweets were labeled by the unsupervised classifier as \textit{defense} or \textit{immigration} rather than \textit{international affairs} as by the supervised classifier. By examining the features associated with each topic in each classifier (see Tables \ref{table:su-features} and \ref{table:un-features}), we can see some of these differences. For instance, the \textit{defense} topic has features ``iran", ``nuclear", and ``strategy" in the unsupervised model and ``veteransday", ``drones", and ``stolenvalor" in the supervised model. In the unsupervised model, topics about \textit{veterans} emerged that were distinct from \textit{defense} and \textit{housing}, where veterans occur in the CAP codebook. So we see that similar terms are associated with different topics in the two types of models, and that explains some of their differences. However, each set of associations is reasonable and interpretable, and that suggests that the models capture different latent properties with their feature-class associations.

Remember that the resulting Cohen's kappa between supervised and unsupervised model results was only 0.262. Common practice suggests that a Cohen's kappa of .21-.40 indicates fair interannotator agreement, with 0 indicating interannotator agreement that approximates random choice \citep{McHugh2012-yb}. In this sense 0.262 does not represent particularly high interannotator agreement between the supervised and unsupervised models. In order to understand why this was, we examined individual cases of disagreement and agreement. We discuss example cases below, how these disagreements contribute to the Cohen's kappa score returned, and what they reveal about the utility of each of the modeling approaches.

Table \ref{table:sample-tweets} describes several sample tweets that were labeled \textit{international affairs} by the supervised model, and indicates how the unsupervised model labeled the same tweet. For example, the unsupervised model labeled the first tweet featured as \textit{agriculture} because of the presence of the hashtag \#FarmBill and mention of sorghum. At the same time, the tweet also reflects international affairs topical focus by mentioning the international trade market and naming other countries in that market. In this case, each of the supervised and unsupervised models captured different, but relevant, thematic focuses of the tweet.

\begin{table*}[h!]
\RawFloats
\begin{tabularx}{\linewidth}{X l l}
\toprule
Tweet Text &  UN1 & UN2\\
\midrule
Sorghum market is very dependent on export. China is main market w/ 75\% of it, then Mexico. \#FarmBill &      \textit{agriculture} &    \textit{uninterpretable}\\
\\
NYC is proud to send 120 \#FirstResponders to help Texans in the wake of \#Harvey https://t.co/vdumEKFnQA &       \textit{emergency response} &    \textit{district affairs}\\
\\
@realDonaldTrump @RepMcEachin @RepJoeKennedy Any time someone is brave enough to serve, we must respect that choice, not stomp on their rights and liberties. \#TransRightsAreHumanRights &	       \textit{civil rights} &    \textit{veterans}\\
\bottomrule
\caption{Sample tweets labeled International Affairs by Supervised Classifier. UN1 refers to the maximum probability class according to the unsupervised classifier, and UN2 indicates the second-highest probability class.}\label{table:sample-tweets}
\end{tabularx}
\end{table*}

The second example where the models diverge was labeled by the unsupervised model as  \textit{emergency response} (highest probability) or \textit{district affairs} (second highest probability). It is possible that the supervised model associated words like ``proud", ``send", and ``help" with the international affairs topic, but the thematic focus of the tweet does indeed appear to be more related to emergency response and district affairs via discussion of ``\#FirstResponders", Hurricane Harvey, and mention of the geographies New York City and Texas. In this case, the unsupervised model captures more relevant themes than the supervised model. These disagreements illustrate that the fixed parameters of the CAP codebook may not capture all types of conversation by MCs on Twitter. Since the unsupervised model represents a bottom-up approach, it is able to capture a new topic not featured in the CAP codebook but that provides additional nuance to the analysis of these tweets.

\subsection{New Insights from the Unsupervised Model} The second example in Table \ref{table:sample-tweets} raises another important difference between our supervised and unsupervised models. Where our supervised model is limited by the policy-focused labels defined by the CAP codebook, our unsupervised model is not. We see that approximately 21.32\%  and 35.76\% of all maximum probability and second maximum probability topics detected fall in topic codes that are not captured in the CAP codebook. These topics (a) reveal that MCs spend a significant portion of their tweets posting about issues other than policy and (b) identify the topics and behaviors beyond policy that they exhibit.

The additional topics the unsupervised classifier identified largely reference activities related to personal relationship building. The ability to distinguish these activities enables analyses of behavior beyond policy debates and on topics such as ``home style". For instance, the topics may make it possible to differentiate Fenno's \citeyearpar{Fenno1978-gd} ``person-to-person" style from ``servicing the district". Relationship-building topics included \textit{sports}, \textit{holidays}, \textit{awards}, \textit{sympathy}, \textit{constituent relations}, and \textit{power relations}. Service-related topics included \textit{district affairs}, \textit{emergency response}, and \textit{legislative process}. Prior research has conducted similar style analysis on smaller sets of tweets around general elections \citep{Evans2014-or}, and our models enable this analysis at scale and throughout the legislative and election calendars.

That somewhere between 20 and 40 percent of all tweets are labeled with these topics tells us that the CAP codebook alone is unable to capture how and why MCs use Twitter. The supervised model does a good job of capturing the policy topics present in the CAP codebook and facilitating the comparative analysis that codebook is designed to support. The unsupervised model enables us to see that MCs invest significant attention in personal relationship building and constituent service on Twitter and to analyze those topics and behaviors that are not clearly policy-oriented.

\section{Recommended Uses of Each Model}

% Given the performance and capabilities of each model, we recommend the supervised model for conducting comparative studies across political systems, and we recommend the unsupervised model for understanding the nuances of the U.S. Congress (e.g., their topics, their non-policy content and behavior). Though useful for comparative studies, our unsupervised model helps us understand that Congress' communications on Twitter do not easily map to the Comparative Agendas Codebook and that there are different ways to classify MCs Twitter behavior that help us understand more about Congress' attention.

The supervised model performs well in assigning tweets topics from an existing measurement system, the CAP codebook, and should be used when researchers wish to do comparative and/or longitudinal analyses. For instance, the supervised model facilitates studies that (a) compare topic attention on Twitter and in speeches \citep{Back2014-et,Greene2017-zm,Quinn2010-nt} or the news \citep{Harder2017-vr} and (b) compare topics in U.S. Congress with topics addressed by other governments \citep[see][for many recent examples]{Baumgartner2019-sx}. Using an existing codebook like CAP facilitates studies that rely on measurement models to study changes over time or to detect differences across contexts \citep{Jones2016-hi}. The shared, established taxonomy, here the CAP codebook, is necessary for these types of analysis.

The supervised model is also better able to capture rare topics. Because the model is trained on all categories, it learns to disambiguate topics such as \textit{housing} and \textit{foreign trade} that the unsupervised model does not detect. These relatively rare categories are significant objects of study despite their infrequency, and therefore the supervised model is a better tool for identifying and analyzing attention to those topics.

The unsupervised model, on the other hand, is most useful for studying the particulars of the U.S. Congress and its Twitter behavior such as communication style and topics unique to the federal and state divisions of authority. The unsupervised model captures the topics and behaviors that MCs exhibit on Twitter that fall outside the purview of the CAP codebook and its related policy studies. For instance, the unsupervised model is a good resource for studying issues such as style \citep{Fenno1978-gd} and framing \citep{Scheufele2007-uj}. The topics this model detects include things such as \textit{constituent service} and \textit{district visit} that can be useful for researchers trying to understand how Congress uses Twitter to communicate with its constituency broadly and not just about clear policy issues. The unsupervised model also allows us to relax the ``mutually exclusive'' criterion of the CAP codebook and to identify the overlap between topics such as \textit{energy} and \textit{environment}. It also reveals the specificity of issues such as \textit{veterans' affairs} that are distributed through the CAP codebook but emerge as one distinct issue area in MCs' tweets. In the following section, we articulate avenues for future research that leverage the models' strengths and weaknesses for different research areas.

% LH working down here May 3

% To further understand the usefulness of each model, we assessed each of their Cohen's Kappa agreement with labels assigned by a human annotator to 13,943 policy-related tweets made available by Russell (as used in \citep{Russell2018-vu,Russell2017-ee}). As expected, since our supervised model was trained on a subset of this data, we found that our supervised model and human-labeled tweets receive a high Cohen's kappa of 0.96, close to nearly complete agreement. The low Cohen's kappa between manual labels and the unsupervised classifier suggest that the taxonomies are mostly independent---these coding approaches measure and capture different aspects of the tweets.

% Our unsupervised model's findings suggest potential extensions to the CAP codebook to reflect tweet-specific U.S. legislative discourse (see codes 24-35 in Table \ref{table:topic-distribution}). It is worth noting that the proposed additional topics largely do not reflect policy topics, but rather different forms of communicating with constituents. These codes describe topics that embody certain communication styles more than topical ideas such as ”Public Lands” or ”Health”. The emergence of these topics suggests that unsupervised topic modeling supports different kinds of analysis than manual and codebook-oriented labeling. Unsupervised topic modeling can also detect communication styles and intents. Further investigation about this potential could yield interesting insights concerning the utility of unsupervised topic modeling to understanding the political communications space.

\section{Future Work}

\subsubsection{Political Attention} One promising set of next steps for research involve using the models to study political attention and social media use in political communication. Our models dramatically reduce the costs of obtaining labeled data for comparative analysis (using the supervised model) and provide a mechanism for identifying additional behavior beyond policy discussions (using the unsupervised model). For example, by using our supervised model, we are able to study the complete 115th U.S. Congress and their relative attention to policies over time. We can then compare attention between subgroups (e.g., parties, chambers, regions of the U.S.) and over time (e.g., around primaries, during recess). When combined with similar advances in using text as data in political science \citep{Grimmer2013-ts}, such as labeling and accessing news content \citep{Saito2018-nc,Gupta2018-mj} and bill text \citep{Adler_undated-yv}, our models also facilitate analysis of the political agenda, enabling researchers to test the paths through which topics reach the mainstream news or enter legislation. The models also facilitate studies of attention-related concepts such as agenda setting and framing \citep{Scheufele2007-uj}. By identifying the parties' rhetoric around a topic, the models make it possible to compare not just whether the parties talk about energy more or less frequently (using the supervised model) but how they discuss it and in combination with what other issues or behaviors (using the unsupervised model).

\subsubsection{International Comparison}
Our supervised model directly allows for comparative analyses by using a standard codebook designed for such studies. Whether our models work on tweets in languages other than English (and marginally Spanish) is an open question. We are currently training models on German parliamentary tweets from 2017 in order to evaluate the potential utility of our approach for understanding politicians' public-facing rhetoric across different languages and country contexts. 

\subsubsection{Evaluating Non-Policy Related Topics} 
The dataset we used to train the supervised model includes additional human-labeled binary tags indicating personal relationship or service-related content in the 113th Congress' tweets. Interestingly, many of these manual tags reflected similar personal relationship or service-related tags that our unsupervised model independently detected. Future work could compare these results---manual labels and unsupervised labels---for this relational content and potentially inform another useful computational tool.

\subsubsection{Model Improvements}
Our models leave some room for improvement, and alternative topic modeling techniques may achieve even more nuanced topic results. %For instance, Moody's lda2vec approach \cite{moody2016} builds upon word2vec's ability to determine word to word relationships in order to augment LDA's ability to determine word to document relationships by building document-level abstractions. Nikita compared the lda2vec's returned topical output with a genism LDA model's returned output (Nikita, 2016). His replication attempt provided a helpful example of comparative model implementation and evaluation methods of both lda2vec and gensim LDA models.

Niu's work with topic2vec modeling explores ways to address the issue of uninterpretable output from LDA models (Niu et al., 2015); he notes that because LDA assigns high probabilities to words occurring frequently, those words with lower frequencies of occurrences are less represented in the derivation of topics, even if they have more distinguishable semantic meaning than the more frequently occurring words. In a similar approach to Moody \citeyearpar{}{moody2016}, Niu tries to combine elements of word2vec and LDA modeling techniques to address the issue of under-specific topics resulting from pure LDA methods. His topic2vec approach yields more specific terms. % (i.e. where LDA yields “patients” and “medical” in each of two topics when run on a particular dataset, topic2vec yields “aricept”, “memantine”, and “enbrel” in one topic and “anesthesiologists”, “anesthesia” and “comatose” in a second topic with the same dataset). 
We did not pursue topic2vec in our tests due to replicability issues encountered with the methodologies proposed, but they offer an interesting avenue to explore in attempts to improve the model.

Leveraging both LDA and community detection via modularity provide an opportunity to compare the types of groupings that emerge from tweet texts in ways that incorporate network structures and relationships \citep{Gerlacheaaq1360}. Efforts to merge topic modeling and community detection also offer additional opportunities to evaluate potentially alternative groupings in an unsupervised way \citep{Mei:2008:TMN:1367497.1367512}. Each of these two approaches to unsupervised text group detection offer interesting opportunities for comparative future work.

\section{Conclusion}
We provide computational models to facilitate research on political attention in social media. The supervised model classifies tweets according to the CAP codebook, enabling comparative analyses across political systems and reducing the labor required to label data according to this common codebook. The unsupervised model labels tweets according to their policy topic, social function, and behavior. It enables nuanced analyses of U.S. Congress, especially the intersection of their policy discussions and relationship building. Together, these two models provide methodological tools for understanding the impact of political speech on Twitter and comparing political attention among groups and over time.

\section{Acknowledgments}
We are especially grateful to Annelise Russell for sharing her data and enabling us to train the supervised classifiers. This material is based upon work supported by the National Science Foundation under Grant No. 1822228.

\bibliographystyle{aaai}
\bibliography{topic-models-for-icwsm.bib}

\end{document}